# Adaptive Normalization in Streaming Data

Vibhuti Gupta
Department of Computer Science
Whitacre College of Engineering
Texas Tech University,
Lubbock, Texas 79415
vibhuti.gupta@ttu.edu

Rattikorn Hewett
Department of Computer Science
Whitacre College of Engineering
Texas Tech University
Lubbock, Texas 79415
rattikorn.hewett@ttu.edu

## ABSTRACT
In today's digital era, data are everywhere from Internet of Things to health care or financial applications. This leads to potentially unbounded ever-growing Big data streams and it needs to be utilized effectively. Data normalization is an important preprocessing technique for data analytics. It helps prevent mismodeling and reduce the complexity inherent in the data especially for data integrated from multiple sources and contexts. Normalization of Big Data stream is challenging because of evolving inconsistencies, time and memory constraints, and non-availability of whole data beforehand. This paper proposes a distributed approach to adaptive normalization for Big data stream. Using sliding windows of fixed size, it provides a simple mechanism to adapt the statistics for normalizing changing data in each window. Implemented on *Apache Storm*, a distributed real-time stream data framework, our approach exploits distributed data processing for efficient normalization. Unlike other existing adaptive approaches that normalize data for a specific use (e.g., classification), ours does not. Moreover, our adaptive mechanism allows flexible controls, via user-specified thresholds, for normalization tradeoffs between time and precision. The paper illustrates our proposed approach along with a few other techniques and experiments on both synthesized and real-world data. The normalized data obtained from our proposed approach, on 160,000 instances of data stream, improves over the baseline by 89% with 0.0041 root-mean-square error compared with the actual data.

## 1. INTRODUCTION
Today's real world data are ample due to various data technologies and heterogeneous sources from Internet of Things to smart cities to health care, social media and financial application. These data do not only increase in volume but also come in high velocity causing potentially unbounded and ever-growing data streams [15]. Data preprocessing is one of the most critical and time consuming steps in knowledge discovery process of data mining [2, 4, 14]. Preprocessing of Big data streams is even more challenging due to inconsistencies of evolving nature, memory and time constraints, limited access of each instance and non-availability of data beforehand [14].

Normalization is a data preprocessing technique that transforms data with attributes of different units into a known uniform scale. This is so that none of the attribute values dominate over others, comparison and aggregation of attributes become easier and the data become better conditioned for convergence [4,15]. Normalization helps prevent skewed results from machine learning algorithms that use a distance measure between attributes [14]. It also improves the efficiency of data analytics. Some of the most well-known normalization methods are min-max, z-score and decimal scaling normalization (see [4]). This paper focuses on Min-Max Normalization for streaming environment.

The above normalization techniques can easily be applied to static data whose properties do not change over time and the entire data are in memory. However, an increasing number of heterogeneous data sources generate evolving data (i.e., continuously change in range values and classes) that are not applicable to existing techniques. The unbounded data streams require normalization techniques that not only adapt well with changes in data but also maintain the time and memory requirements.

Much research on adaptive learning in streaming environment [1, 5, 12, 16] deals with learning models that adapt with evolving data distribution. However, preprocessing data parameters are usually kept fixed or manually adjusted before learning, which can lead to poor prediction results [18]. Adaptive normalization is rarely explored for Big Data streams in the research literature. Majority of studies (e.g., [8, 11, 13, 18]) in adaptive normalization are with non-stationary (i.e. time series) finite datasets (i.e. entire data in memory) where their evaluation ties with specific purpose (e.g., classification) using various learning models. Work in [10] proposes an adaptive normalization technique for network monitoring. However, none of these approaches address normalization of evolving Big data streams, which demands a distributed approach.

This paper proposes a distributed approach to adaptive normalization for Big data stream. Using sliding windows of fixed size, it provides a simple mechanism to adapt the statistics for normalizing changing data in each window. Implemented on *Apache Storm*, a distributed real-time stream data framework, our approach exploits distributed data processing for efficient normalization. Unlike other existing adaptive approaches that normalize data for a specific use (e.g., classification), ours does not. Moreover, our adaptive mechanism allows flexible controls, via user-specified thresholds, for normalization tradeoffs between time and precision. The paper illustrates our proposed approach along with a few other techniques and experiments on both synthesized and real-world data on electricity pricing over different periods of time in one day. The normalization is needed to obtain consistent scales.

The rest of the paper is organized as follows: Section II discusses related work. Section III describes our approach followed by illustrative example and experimental results in Section IV and V. Section VI concludes the paper.

## 2. RELATED WORK

Recent research on Adaptive Normalization methods includes sliding window techniques (e.g., [7, 9]), adaptive methods for time series forecasting [11, 13], and those that are designed for data stream mining [8, 10].

Most of the adaptative normalization techniques with sliding windows [7, 9] employ a fixed window size to incrementally normalize data one window at a time. These techniques are similar to ours in that both use sliding windows of a constant size. Unlike these approaches whose statistics (e.g., min, max values) for normalization change for each window, ours do not and only change when the degree of deviation of the new statistics from the old ones is high enough (more details later).

The adaptive normalization techniques in [11, 13] are used for transforming the non-stationary time series sequence into the normalized data that help improve construction of learning models. In particular, exponential moving average is applied to non-stationary time series sequence that are transformed into a stationary sequence by dividing the data into disjoint sliding windows [11]. Further, outliers are removed and then all disjoint sliding windows are explored to obtain global min-max for normalization. Here the adaptiveness is in the transformation of non-stationary data to stationary. Another adaptive normalization technique applies shifting and scaling to the data and adaptively change the normalization scheme with the changes in the data [13]. This technique is a normalization layer for applying deep learning model to non-stationary time series. Our approach is different because it does not aim to improve learning models.

The adaptive normalization techniques in [8, 10] are closest to our work in spirits. First, like ours, the preprocessing is decoupled from learning. Second, it aims to prevent the unnecessary adaptive normalization without significant changes in the data stream. To do this, two metrics are used. One is to identify the noise/outliers and the other is to signal significant changes in the stream [8]. If both are identified then the statistics for normalization are changed. Since values that are not in current window range are considered as noise and outliers (which is not necessary the case), normalizing them with old min-max values without adapting them will lead to error. Therefore, these techniques work well for sharp/abrupt changes but not for gradual changes. Unlike the above techniques, our more conservative adaptive mechanism allows normalization of gradual changes as well as abrupt changes.

In addition, overall none of [8, 11, 13] considers the applicability of their adaptive approaches to Big Data streams whereas our approach does by providing an adaptive normalization implemented on a distributed framework.

## 3. NORMALIZATION TECHNIQUES

In this section we briefly describe our early adaptive normalization approaches using fixed-size sliding window (Section 3.1) followed by proposed normalization approach in detail (Section 3.2).

### 3.1 Our Early Adaptive Normalization

In this section, we will describe earlier normalization approaches in an increasing order of their adaptive evolution. These approaches fall into the following four categories:

1. Known Value Range Normalization
2. Window based non-adaptive Normalization
3. Per window Normalization
4. Adaptive Normalization with significant changes.

All the above approaches are referred by their number throughout this paper. Known value range Normalization approach (i.e. approach 1), assumes that the min-max values are known before the normalization and fixed throughout the data stream. Data in each window is normalized with these fixed parameters. Even though the parameters are fixed, the normalized values with this approach are considered as ground truth for our experiments since the actual min-max values are known for the whole data. Thus, normalized values are considered correct and the approach as baseline in this paper.

For the rest of the approaches (i.e. 2,3,4), it is assumed that the actual min-max values are unknown initially since access to whole data is not available during analysis. In Window based non-adaptive Normalization approach (i.e. approach 2), data parameters of first window are used to normalize the whole data stream. In this approach instead of adapting the statistics with changes in the stream, fixed statistics of first window is used which leads to out of bound error. Hence, non-adaptive approach. In per window normalization approach (i.e. approach 3), each window data points are normalized with their respective min-max values which is similar to [7][9]. Statistics are updated for each new window. Hence, this approach works well for significant changes but leads to out of bound error for non-significant changes. It can be useful for some special cases of data with abrupt changes.

Adaptive Normalization with significant changes (i.e. approach 4) approach adapts the statistics (i.e. min-max) with significant changes. This approach is similar to approach [8] in the sense that for the non-significant changes, old min-max values of are applied to normalize current window elements. Hence, normalization is dependent on the statistics of first window for every sequence of data points with non-significant changes in the data stream which leads to out of bounds error.

Approach 5 is our proposed approach that handles the issues with all the above approaches Next section explains how we processed the whole data stream with fixed-size sliding windows followed by our proposed normalization approach in detail.

### 3.2 Proposed Approach

Initially the input data points from data stream are collected into a fixed-size sliding window until the number of points are equal to the predefined size of window. Then, the required statistics are captured and used to normalize the data points in the window. After that window slides to process further data points. This entire process continues until all the data points are normalized from the data stream.

Our adaptive normalization approach starts by computing the required statistics (i.e. min-max, mean) from the first window to set the initial reference parameters for Normalization. Data points in first window are normalized by these reference parameters. Then we check for significant changes in the data stream to determine the point of adaptation. Significant changes indicate that the data distribution has changed. Thus, data parameters needs to adapt in order to normalize in the new environment. To detect the changes, we compute the percentage change in mean values between the current and previous window and if it exceeds the threshold, then old reference min-max values are replaced by new reference min-max. When the percentage change in mean values does not exceed the threshold, then the change is considered as non-significant but to maintain the actual range and remove the dependency on one window, either of reference min/max are incrementally updated. In our approach we have used percentage change in mean values as a change detector since it accurately captures the changes for numeric data streams. We assume that there are no extreme (i.e. very high/very low values) outliers in the data.

Figure 1 shows the *Adaptive-Normalization* algorithm taking *S*, numeric data stream, as input to generate $S_n$ which is normalized data stream.

---

**Algorithm** *Adaptive-Normalization algorithm*

---

**Inputs:** *S*, Numeric data stream; *N*, Window size representing number of data points; *i*, window number; $\delta$, Threshold for % change in mean values between current and previous window elements
**Output:** $S_n$, Normalized Data Stream
01: **if** $i==1$ //First window of size *N*
02: $refmin \leftarrow min_i$ ; $refmax \leftarrow max_i$
03: Normalize $W_i$ elements with *refmin* and *refmax*
04: **else if** $i > 1$ **then** // Other windows of size *N*
05: *Repeat*
06: $cur\text{-}window \leftarrow W_i$; $prev\text{-}window \leftarrow W_{i-1}$
07: **if** $((|\ mean_i - mean_{i-1}\ |)\ /\ mean_{i-1}) \geq \delta$
//Update *refmin* and *refmax* values
08: $refmin \leftarrow min_i$ ; $refmax \leftarrow max_i$
09: Normalize $W_i$ elements with *refmin* and *refmax*
10: **else** //Non-significant change
// compute current min and max values
11: $currmin \leftarrow min_i$ ; $currmax \leftarrow max_i$
//Update reference min and max values
12: $refmin \leftarrow MIN(currmin,refmin)$ //Update min value
13: $refmax \leftarrow MAX(currmax,refmax)$ // Update max value
14: Normalize $W_i$ elements with *refmin* and *refmax*
15: **end if**
16: $i \leftarrow i+1$
17: *Until* there are no data points in stream *S*
18: **end if**
19: **return** $S_n$

**Figure 1. Adaptive Normalization**

Data points are collected until there are *N* data points in the window. Initial min-max values are computed to set reference min-max and used to normalize first window elements (Lines 01-03). As the stream progress, window slides to process next *N* data points. Current window is represented by *cur-window* and previous window by *prev-window* (window before current one),$W_i$, represents i-th window. Now the percentage change in mean values are computed (Line 07), $mean_i$ represents the mean of i-th window(i.e. current), $mean_{i-1}$ is the mean of (i-1) th window (i.e. previous), if it exceeds threshold then reference min-max values are updated with new min-max (Lines 08-09), otherwise reference min-max values are incrementally updated with every new window and used to normalize current window elements (Lines 11-14). This entire process repeats until all data points are normalized (Line 05-17).

The proposed approach has been implemented in *Apache Storm* [19], a distributed real-time processing framework for Big Data Streams. Storm provides two types of windowing support i.e. sliding windows and tumbling windows. Sliding windows slide with every new incoming data point/sliding interval but tumbling windows slide only with fixed number/interval of data points. We have used tumbling windows in this paper. Storm processes data in real time using *Spouts* and *bolts* as components to form *topology* that will be submitted to cluster to perform distributed processing. A detailed description about Storm can be found at [19].

# 4. ILLUSTRATION

This section illustrates all approaches (i.e. 1,2,3,4,5) with an example. Suppose we have a numeric data stream that has varying range values and we need to normalize it using fixed size sliding window. The input data is having range [20-40] and [80-100], window size is 5. Figure 2 shows the actual data with 2 sliding windows (A&B), we will show the normalized versions of these 2 windows further in different cases of adaptiveness.

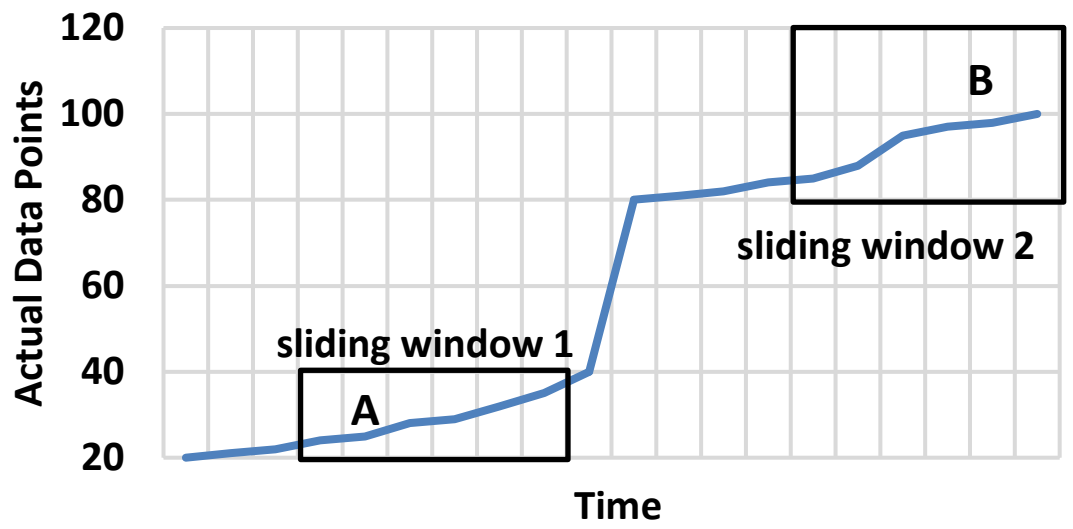

**Figure 2. Actual Data with 2 sliding windows**

Figure 3 shows normalization with method 1 for both windows. As we can see window A & B both normalized correctly since actual min-max values are known.

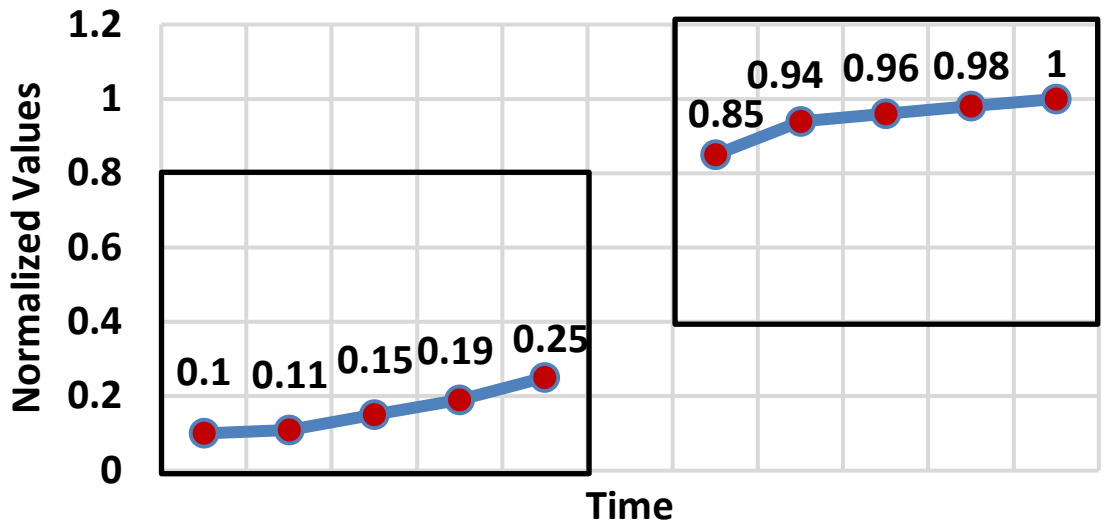

**Figure 3. Method 1 applied to Fig 2 boxes A & B**

Figure 4 shows method 2 , in this case window A normalizes correctly since first window statistics are used to normalize but window B doesn't due to no adaptation, which leads to out of bound normalized values. Figure 5 shows that the statistics are updated for each window which produces normalized values closer to baseline but it adapts regardless of degree of deviation in the statistics which is not necessarily the case every time. Figure 6 shows the normalization with method 4, as we can see that both windows have out of bounds normalized value. Here the min-max values of previous window are directly applied when there are non-significant changes similar to approach in [8].

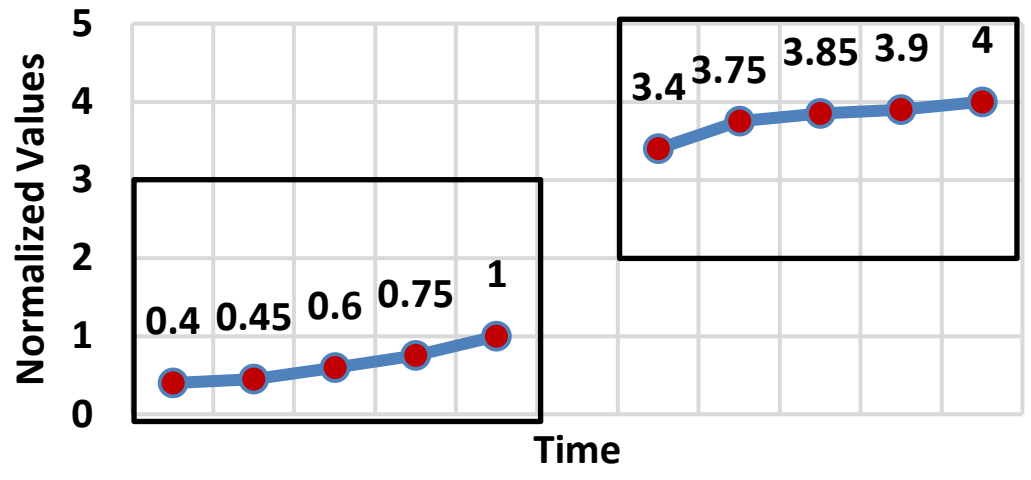

**Figure 4. Method 2 applied to Fig 2 Boxes A & B**

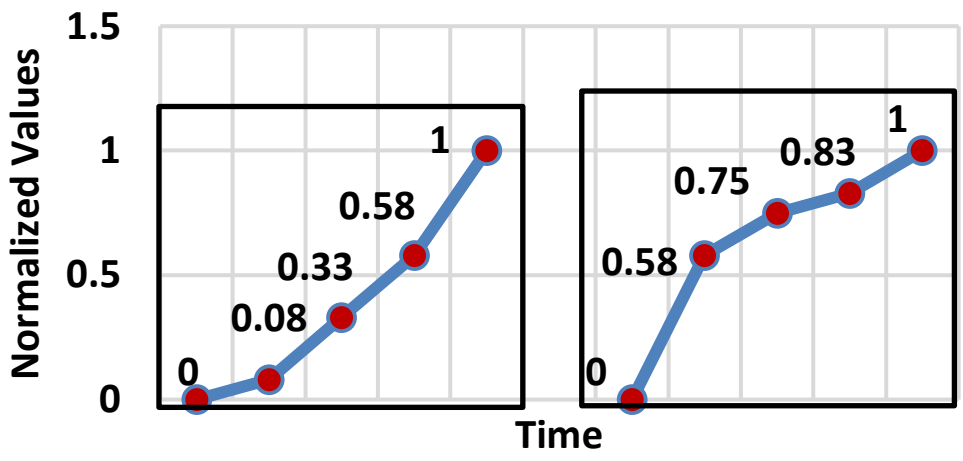

**Figure 5. Method 3 applied to Fig 2 Boxes A & B**

So even though there are non-significant changes we have to update the min/max values incrementally to maintain the range.

As shown in Figure 4, window 1 normalizes correctly due to this reason. So while normalizing data stream with fixed size window we have to keep track of data parameters with significant as well as non-significant changes to avoid out of bounds error. Figure 7 shows our proposed approach i.e. method 5. As shown in figure 2, there is an acceptable degree of deviation where min-max values changes from [20-40] to [80-100], at this point the min-max values are updated in second window as shown in figure 7 which shows that our approach adapts with the changes in data .

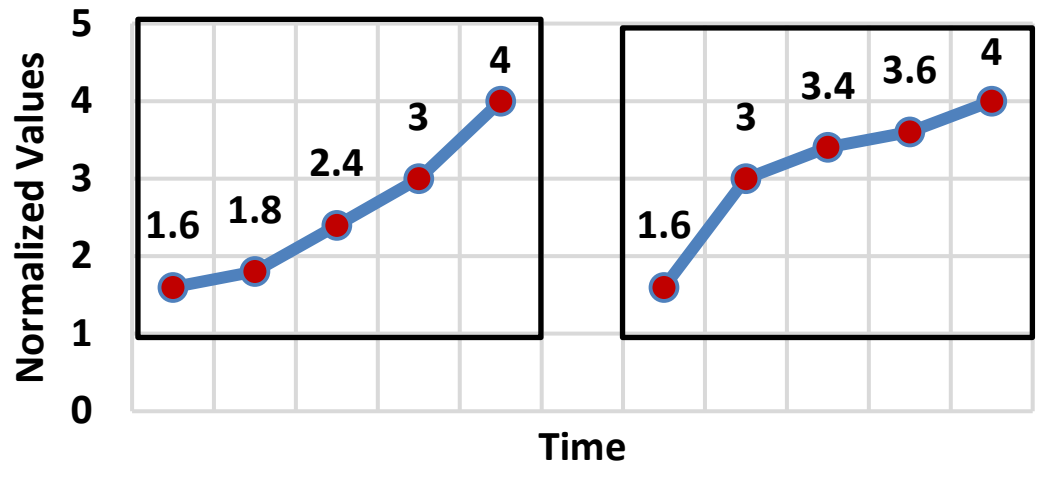

**Figure 6. Method 4 applied to Fig 2 boxes A & B**

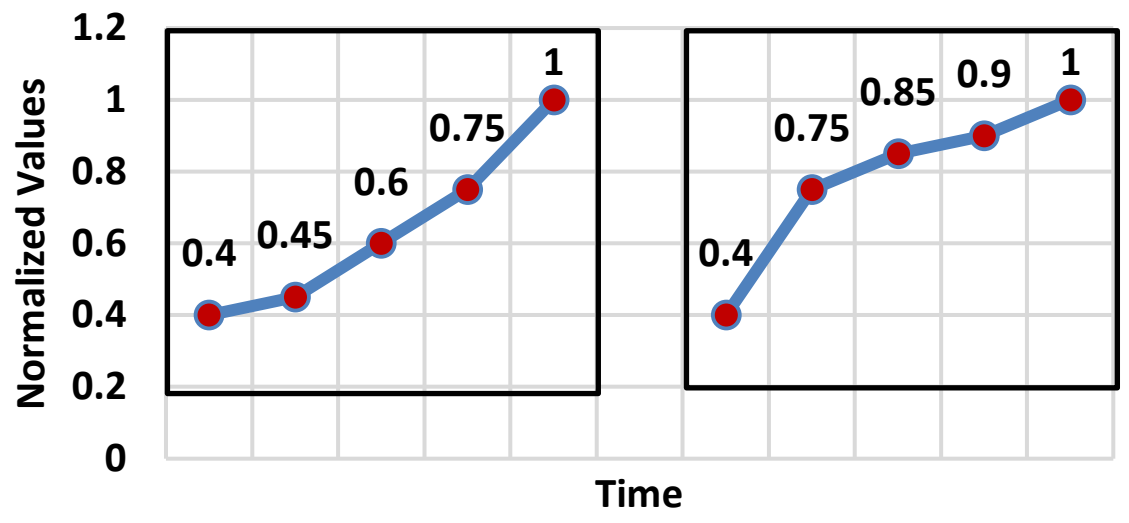

**Figure 7. Method 5 applied to Fig 2 boxes A & B**

# 5. EXPERIMENTS AND RESULTS

This section provides the experimental results to evaluate the effectiveness of our proposed approach in Big Data Stream Infrastructure.

## 5.1 Dataset & Experimental Setup

We applied our approach to numeric one dimensional synthetic dataset and real world Electricity Market (EM) dataset. Synthetic dataset is generated within the range of [1-60] and divided into 4 subranges i.e. [1-5], [1-10], [30-50] and [30-60] to evaluate the effectiveness of our approach during significant and non-significant changes. It is generated with varying sizes i.e. 20k, 40k, 80k, 160k with each of them divided into 4 subranges equally. Electricity Market (EM) dataset [20] is well known for stream mining research, it contains 45312 instances of electricity prices drawn from 7 May 1996 to 5 Dec 1998 with one instance for each half hour. It has seven dimensions with five of them numerical and rest two are date and time values. We removed the date and time values to make the dataset fully numerical. And then applied our approach to normalize varying electricity prices.

We have done experiments in *Storm framework* in single and multiple processors to evaluate the effectiveness of our approach in a Big Data Stream environment. The Storm cluster is composed by a varying number of virtual machines (VMs or processors) (i.e., 1, 2, 4, and 8) in a system with Intel Core -i7-8550U CPU 2 GHz processor, 16 GB RAM 8 cores and 1TB of Hard disk. Each virtual machine is configured with 4 vCPU and 4 GB RAM with Ubuntu 14.04.05 64 bits OS along with the JDK/JRE v 1.8. The *Apache Storm* version used is 1.1.1 with *zookeeper* 3.4.9 [19].

## 5.2 Experimental Results

This section provides the experimental results with single and multiple processors. We used Root Mean Squared Error (RMSE) as an evaluation metric to measure absolute performance of our proposed approach. Table 1 compares the average RMSE values of each method with the baseline for synthetic and real datasets. The size of synthetic dataset is 160000 while real world data 45312. As shown in table 1, our proposed approach 1 vs. 5 outperforms other approaches for both datasets. Our proposed approach shows an improvement in root mean square error with 89% for Synthetic and 88.4% for real world data. Hence best performance among other approaches. Approach 2 performs worst as compared to others since it's non-adaptive and has fixed data parameters throughout the data stream.

**Table 1. Comparison of root mean square error results**

| | Comparisons | | | |
|---|---|---|---|---|
| **Dataset** | **1 vs. 2** | **1 vs. 3** | **1 vs. 4** | **1 vs. 5** |
| Synthetic | 3.93% | 0.46% | 0.54% | **0.41%** |
| Utility | 3.01% | 0.58% | 0.65% | **0.35%** |

Approach 4 has higher error value than 3 since it adapts well with significant changes but has fixed parameters with non-significant changes as shown in Figure 6. Approach 3 has lower error value since it adapts with every window but it is not beneficial every time.

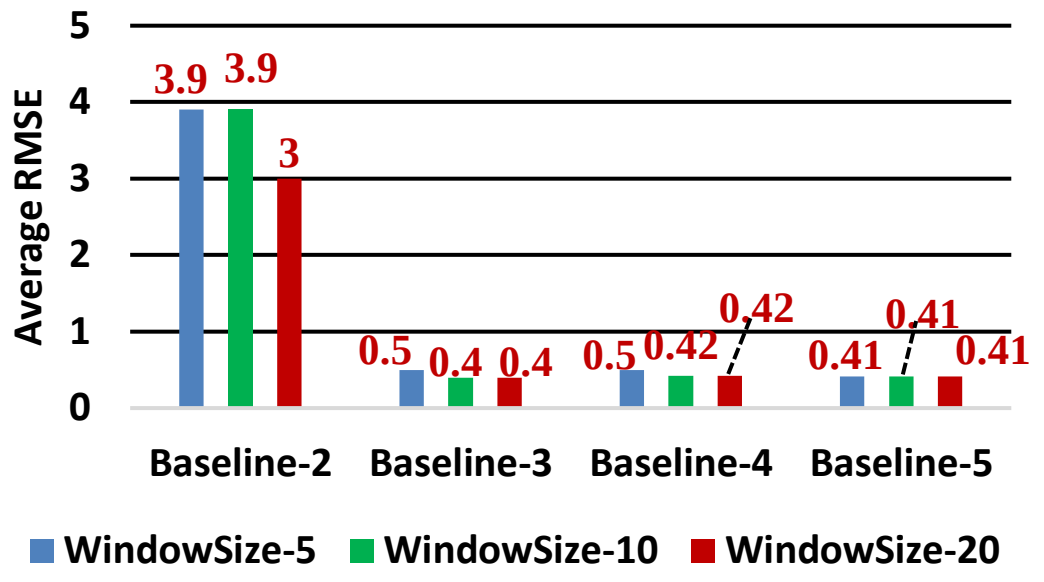

**Figure 8. Average RMSE on various sizes**

Figure 8 shows the variation of average RMSE with different window sizes for synthetic data. It decreases with increasing window size for all approaches because as the window size increases, determining the correct min-max parameters become easier which leads to more accurate values. One interesting aspect is that error remains constant for our approach since we adapt with significant changes and maintain range with non-significant changes. This shows that we can keep any window size to apply our approach.

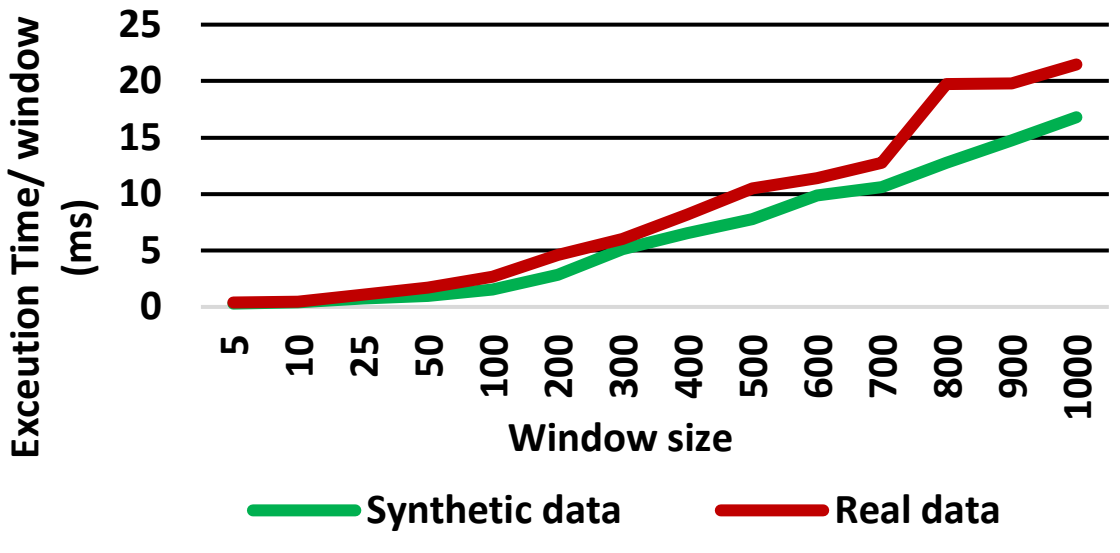

**Figure 9. Execution time per window**

Figure 9 shows the execution time per window with varying window sizes for both synthetic and real data for our proposed approach. It shows that the processing time per window is proportional to window sizes since larger the window size, more time it will take to determine the statistics, hence more time for normalization. As shown in Figure 9, Both the datasets have almost the same execution time for smaller window sizes (smaller than 50). However, for larger window sizes (greater than 50), execution time for the synthetic data is less than real data since there are more frequent changes in real data stream as compared to synthetic one due to which statistics (i.e. min-max) has to be determined and updated more frequently which increases the time to normalize each window .

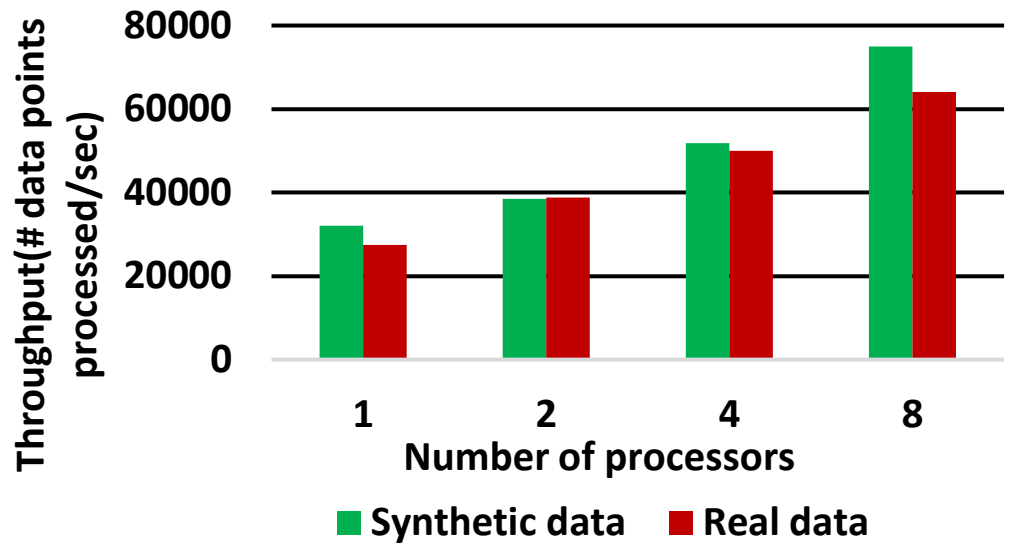

**Figure 10. Throughput with more processors**

Figure 10 compares the throughput for proposed approach as the number of processors increase. Throughput is the total number of data points processed per unit time (i.e. seconds in our case). We ran the experiment for a session of 5 minutes in each of the case (i.e. with no. of processors as 1,2,4,8). As shown in figure 10, throughput improves from 32041 points/sec to 75010 points/sec for synthetic data and from 27461 to 64044 points/sec for real data as we increase from a single processor to eight processors.

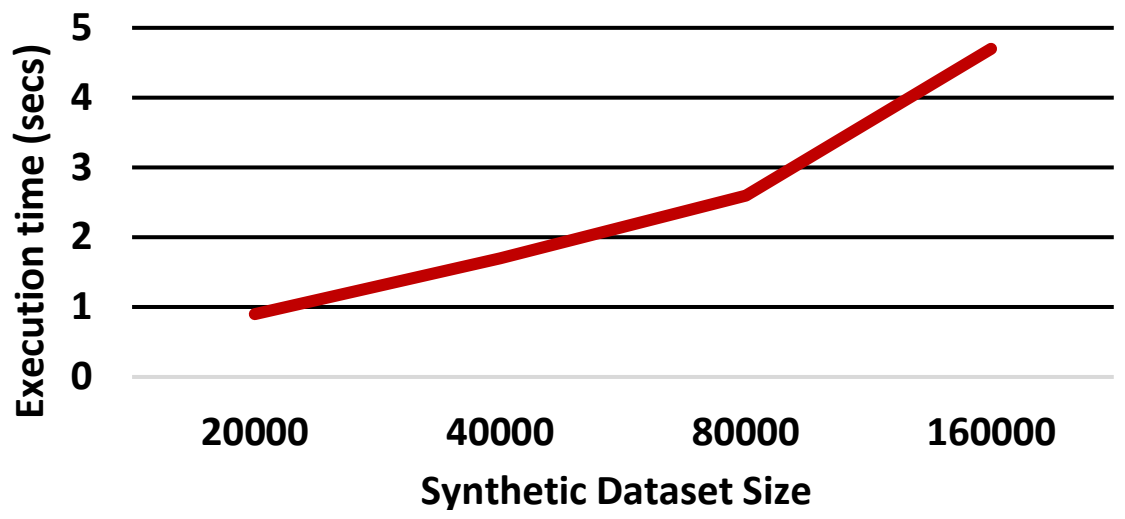

**Figure 11. Execution time with varying dataset size**

The throughput shows a slight increase as we doubled and quadruples the number of processors but reached highest of 134% increase with synthetic and 133% with real data when the number of processors are eight. The number of processed data points depend upon the speed of execution. As we increase the amount of parallelism by adding more processors, the rate of processing data points increases resulting in the improvement of the throughput.

We also experimented with the scalability of algorithm with the size of dataset. This experiment is performed with synthetic dataset. As shown in figure 11, execution time increases almost linearly with increase in dataset size. It is almost doubled from 0.9 to 1.7 secs as the dataset size doubled from 20000 to 40000 data points and reached highest (i.e. 4.7 secs) with 160000 data points. This shows that our approach performs well as expected Thus, the proposed approach provides an efficient Normalization technique that has promising applicability to Big Data streams.

# 6. CONCLUSION

This paper presents a distributed approach for adaptive and efficient Normalization for Big Data Stream. The approach is general in the sense that it can be applied to any Numerical data stream in any domain. The experimental results show that the proposed approach is scalable and can be applied with any fixed-size window. Future work includes experiments on different domains and application of this approach for diverse datasets to reduce complexity and prevent mismodeling. Additional research using different change detection techniques and provide dynamic adjustment of threshold are required to improve the normalization techniques.

# REFERENCES

[1] Elwell, R., & Polikar, R., "Incremental learning of concept drift in nonstationary environments", *IEEE Transactions on Neural Networks*, pp. 1517-1531,2011.

[2] García, S., et al., "Tutorial on practical tips of the most influential data preprocessing algorithms in data mining", *Knowledge-Based Systems*, 98, 1-29, 2016.

[3] García, S., et al., "Big data preprocessing: methods and prospects", *Big Data Analytics*, 1(1), 9, 2016.

[4] García, S., et al., "Data Preprocessing in Data Mining", *Springer*, 2015.

[5] Gu, X. F., et al., "An improving online accuracy updated ensemble method in learning from evolving data streams", *In Proceedings of 11th International Computer Conference on Wavelet Actiev Media Technology and Information Processing (ICCWAMTIP)* ,pp.430-433, 2014.

[6] Han, J., et al., "Data mining: concepts and techniques", San Francisco*: Morgan Kauffman*,2001.

[7] Haykin, S., et al., "Neural networks and learning machines" , *Upper Saddle River: Pearson education*,2009.

[8] Hu, H., & Kantardzic, M. ,"Smart preprocessing improves data stream mining", *In Proceedings of 49th Hawaii International Conference on System Sciences (HICSS)*, pp. 1749-1757, 2016.

[9] Lin, J., & Keogh, E., "Finding or not finding rules in time series", *In Applications of Artificial Intelligence in Finance and Economics*, pp.175-201, Emerald Group Publishing Limited,2004.

[10] Lopez, M. A., et al., "A fast unsupervised preprocessing method for network monitoring", *Annals of Telecommunications*, 74(3-4), 139-155, 2019.

[11] Ogasawara, E., et al., "Adaptive normalization: A novel data normalization approach for non-stationary time series", *In Proceedings of International Joint Conference on Neural Networks (IJCNN)* , pp. 1-8, 2010.

[12] Parker, B. S., et al., "Incremental ensemble classifier addressing non-stationary fast data streams", *In Proceedings of IEEE International Conference on Data Mining Workshop* , pp. 716-723, 2014.

[13] Passalis, N., et al. "Deep Adaptive Input Normalization for Price Forecasting using Limit Order Book Data." arXiv:190.07892, 2019.

[14] Pyle, D., Data preparation for data mining, *morgan kaufmann*,1999.

[15] Ramírez-Gallego, et al., "A survey on data preprocessing for data stream mining: Current status and future directions", *Neurocomputing*, 239, 39-57, 2017.

[16] Street, W. N., & Kim, Y., "A streaming ensemble algorithm (SEA) for large-scale classification", *In Proceedings of the seventh ACM SIGKDD international conference on Knowledge discovery and data mining* , pp. 377-382, 2001.

[17] Tan, P. N., et al., "Association analysis: basic concepts and algorithms", *In Introduction to Data mining (Vol. 321321367). Boston, MA: Addison-Wesley*, 2005.

[18] Zliobaite, I., & Gabrys, B., "Adaptive preprocessing for streaming data", *IEEE transactions on knowledge and data Engineering*, 26(2), 309-321, 2012.

[19] Toshniwal, Ankit, et al. ,"Storm@ twitter," *In Proceedings of the ACM SIGMOD international conference on Management of data*, ACM, 2014.

[20] Harries, M., & Wales, N. S., *Splice-2 comparative evaluation: Electricity pricing*, 1999.